\documentclass[aps,prl,twocolumn,superscriptaddress,floatfix,nofootinbib]{revtex4-2}

\usepackage{graphicx}
\usepackage{amsmath,amssymb}
\usepackage{bm}
\usepackage{xcolor}
\usepackage{booktabs}
\usepackage[colorlinks=true,linkcolor=blue!55!black,citecolor=blue!50!black,
 urlcolor=blue!55!black]{hyperref}

\begin{document}

\title{Emergence of an acoustic black hole in a lattice model}

\author{Antonino Flachi}
\affiliation{Department of Physics \& Research and Education Center for Natural
Sciences, Keio University, Yokohama, Japan}

\date{\today}

\begin{abstract}
We build an acoustic black hole in a lattice model: a draining vortex evolved by the local collide-and-stream rule of a lattice-Boltzmann model, with no acoustic metric imposed, the geometry being measured from the steady state. A horizon emerges, first located from the measured Mach profile and then confirmed by phonon ray tracing on the measured metric, and the ringdown after a perturbation follows approximately the quasinormal frequencies of the draining bathtub. The residual departures are traced to the cell-scale structure of the horizon, which is not an ideal absorber. We model it in two ways: as a reflective horizon on the ideal bathtub, with two free parameters (the modulus and phase of the reflectivity), and as the sonic point of the realised flow absorbing in its own frame, with no adjustable parameter. The second describes the horizon and the non-rotating spectrum best. Superradiance has two ingredients: the emergent geometry, which amplifies in a loss-free calculation on the measured metric, and the horizon, whose absorption frame sets the width and strength of the amplified band but leaves no trace in the ringdown. Both models are used to examine superradiance, which has not been observed in the present toy model: whether the lattice hole superradiates dynamically remains open. The lattice leaves its own imprint as well: a fourfold anisotropy of the sonic surface and a surface gravity set by the cell-scale structure of the horizon.
\end{abstract}

\maketitle

\paragraph{Introduction.} The analogue-gravity program~\cite{Unruh1981,Visser1998,BarceloLR,Schutzhold1998,Visser2005} rests on the observation that sound waves on a moving fluid background obey the same wave equation as a massless scalar field on a curved Lorentzian spacetime, $\Box_g\phi=0$, with an acoustic metric $g_{\mu\nu}$ built from the local density $\rho$, flow $\mathbf v$, and sound speed $c_s$. Where the normal flow speed crosses $c_s$, an effective event horizon forms: phonons inside cannot propagate back out. This correspondence has driven a generation of experiments: in water tanks~\cite{Weinfurtner2011,Torres2017}, Bose--Einstein condensates~\cite{Steinhauer2016}, and optical fibres~\cite{Drori2019}.
In analogue models it is the theoretical formulation, i.e. the emergent ``acoustic'' metric and the wave equation, that provides the substrate for the analogy and the framework for comparison with experiments.

Here we build such an analogue black hole in a lattice fluid. A draining-bathtub flow is evolved with the lattice Boltzmann (LB) method~\cite{Succi2001,Chopard1998}:
the simulation is defined entirely by its microscopic update rule, together with the boundary conditions and a static radial body force acting on the momentum balance; no acoustic metric enters at any point, and the compressible hydrodynamics, and with it the acoustic geometry, emerge at scales large compared to the lattice spacing~\cite{FHP1986,Chopard1998}. A stable horizon is not automatic, since not every arrangement of sub- and supersonic flow yields one. We treat the output as experimental data: no metric is prescribed, and every quantity compared with the analogue-gravity framework~\cite{Visser1998,CLY2004}, the horizon location, the quasinormal frequencies, the realised metric, is \emph{measured} from the running simulation, as in a laboratory. The lattice model thus sits midway between theoretical analogues and their experimental realisations.

The lattice character of the model is instrumental to this perspective. The LB method
retains the defining architecture of a cellular automaton~\cite{Wolfram1983,FHP1986,LGCA,Toffoli,Chopard1998}: \emph{discrete space}, a regular, fixed lattice; \emph{discrete time}, synchronous distinct steps; and \emph{strictly local rules}, since the collision step occurs within a single cell and the streaming step only transfers data to immediate neighbours. Viewed this way, the lattice model comes with its own built-in short-distance physics. The square lattice breaks rotation invariance at the cell scale, an anisotropy that surfaces precisely where the horizon probes short distances (see below); the discreteness provides a physical ultraviolet cutoff, the counterpart of the capillary or healing scale of laboratory analogues; and, taken literally, such models are a concrete toy example of a discrete substrate underlying an emergent relativistic phenomenology. We return to these points in the Discussion.

This places the model differently from earlier lattice treatments of horizons. Lattice field theories on a prescribed black-hole background were used to study the Hawking process with a strict short-distance cutoff~\cite{CorleyJacobson1998,JacobsonMattingly2000}, and quantum walks reproduce Dirac fermions on curved backgrounds, including Schwarzschild~\cite{DiMolfetta2013}; in all of these the metric is an input. The LB method has been extended to relativistic hydrodynamics and to flows on curved manifolds~\cite{Mendoza2010,Mendoza2013}, again with the geometry prescribed. To our knowledge, no cellular-automaton, lattice-gas or LB realisation of an analogue black hole has been reported; here the geometry is an output. At the other end, 't Hooft's programme~\cite{tHooft2016} takes deterministic cellular automata, from which LB models descend, as candidate substrates for quantum mechanics itself. We make no such claim, but a lattice fluid that produces a horizon and a ringdown from local deterministic rules is a concrete instance of the emergence that programme presupposes, and shows which continuum features survive and which become substrate-specific.

\paragraph{The lattice model.} The model we adopt is the standard D2Q9 LB scheme~\cite{Chopard1998,Lallemand2000} (details in Appendix~\ref{app:sec:lbm}): a square lattice with each cell holding nine real numbers $f_i$, the populations moving along the nine discrete velocities $\mathbf e_i$ (rest, four axial, four diagonal), with weights $w_i$ and lattice sound speed $c_s=1/\sqrt3$, fixed by the lattice geometry. The entire dynamics is the two-line collide-and-stream rule
\begin{align}
\mathsf{collide:}\quad & f_i^\star=f_i - \big[\mathsf M^{-1}\mathsf S\,\mathsf M\,(f-f^{\rm eq})\big]_i, \label{colliderule}\\
\mathsf{stream:}\quad & f_i(\mathbf x+\mathbf e_i,\,t{+}1)=f_i^\star(\mathbf x,t), \label{streamrule}
\end{align}
where the collision relaxes the populations of a cell towards the local equilibrium
\begin{equation}
f_i^{\rm eq}
=w_i\,\rho\Big[\,1
+3(\mathbf e_i\!\cdot\!\mathbf u)
+\tfrac92(\mathbf e_i\!\cdot\!\mathbf u)^2
-\tfrac32|\mathbf u|^2\,\Big],
\end{equation}
and the streaming moves each population to the adjacent cell in its own direction. The density and momentum are the two moments
\begin{equation}
\rho=\sum_i f_i,\qquad \rho\,\mathbf u=\sum_i \mathbf e_i\,f_i,
\end{equation}
conserved exactly by the collision. The matrix $\mathsf M$ maps the $f_i$ onto mass, momentum, energy and stress moments and $\mathsf S$ holds their relaxation rates (explicit forms in Appendix~\ref{app:sec:lbm}); the shear rate $s_\nu=1/\tau$, with $\tau=0.70$, sets the viscosity $\nu=c_s^2\big(1/s_\nu-\tfrac12\big)$, and the ghost rate is tied to it by the rule of Ref.~\cite{Ginzburg2008}, $\left({1}/{s_\nu}-1/2\right)\left({1}/{s_q}-1/2\right) =1/4$, which, in our runs, silences spurious reflections at the boundaries.

\begin{figure*}[t]
\centering
\includegraphics[width=\textwidth]{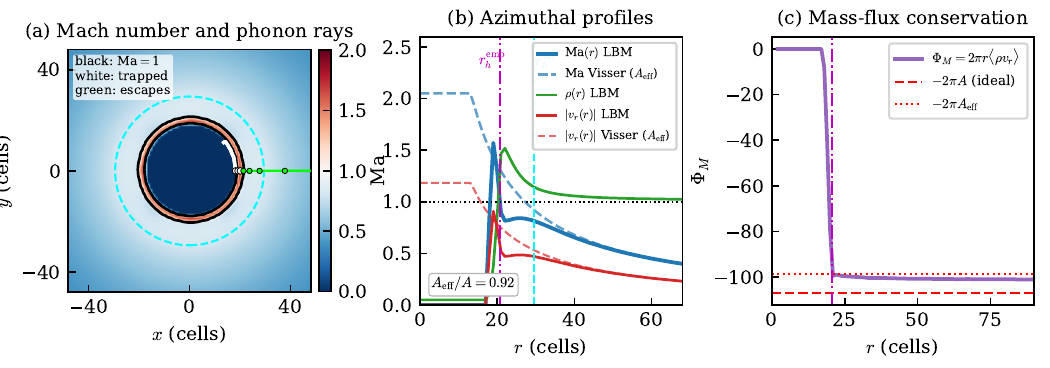}
\caption{\label{fig:diag}
The lattice acoustic black hole, diagnosed from its steady state
($A=17$, $B=0$, $448^2$ cells, body-force core; $r_h^{\rm emp}=20.7$).
(a) Mach number $|\mathbf u|/c_s$ (at $B=0$ the radial Mach number); black contours
mark $\mathrm{Ma}=1$: the outer contour is the emergent sonic horizon and the
inner one the deceleration edge of the central absorber, the two bounding the
supersonic annulus; the dashed cyan circle is the ideal
$r_h=A/c_s$. Overlaid are phonon rays traced on the measured acoustic metric: rays launched inside the
sonic surface (launch radii $18.3$--$19.9$) are trapped (white), those outside
($21.1$--$55.5$) escape (green). (b) Azimuthally averaged Mach, density and radial-velocity
profiles versus the Visser form \cite{Visser1998} at the realised $A_{\rm eff}$ (dashed;
the ideal density, $\rho=1$, coincides with the dotted unity line);
$r_h^{\rm emp}$ (magenta) and the ideal $r_h^{\rm an}=A/c_s$ (cyan). (c) Radial
mass flux $\Phi_M(r)$: a flat plateau (steady-state continuity) at
$-2\pi A_{\rm eff}$, below the ideal $-2\pi A$ by the realised-drain factor.}
\end{figure*}

\paragraph{Emergent horizon, diagnostics and ray tracing.} The draining-bathtub geometry, $\mathbf v_{\rm
db}=(-A\hat{\mathbf r}+B\hat{\bm\theta})/r$, can be realized by imposing a specific \emph{boundary} behavior with the outer rim
reset at each tick to $f^{\rm eq}(\rho{=}1,\mathbf v_{\rm
db})$ and a soft central sink
to $f^{\rm eq}(\rho_{\rm sink},0)$.
If no body force acts in the bulk, the transonic flow and its horizon are an
emergent steady solution.
In this case,
the sonic surface forms but it remains pinned to the soft-drain edge.
A stable horizon in the bulk can be realized by adding a \emph{body-force core}, i.e., a radial force $g_r=-(A^2+B^2)/r^3$ that makes $\rho=1$, $v_r=-A/r$ an \emph{exact} steady solution in the continuum; this places a well-resolved horizon in the bulk for any chosen $A$. The body-force core is introduced in the model following Ref.~\cite{Guo2002} (see Appendix~\ref{app:sec:lbm}).
This is the construction used for Fig.~\ref{fig:diag}. The flows reported here converge to a steady state and every measurement is taken on that state.
In the steady state, the flow is transonic: the surface where the radial Mach
number reaches unity is the sonic horizon, predicted at $r_h=A/c_s$, with
surface gravity $\kappa=c_s^3/A$ (in the convention of Ref.~\cite{Visser1998}, $\kappa=c_s\,|dv_r/dr|_{r_h}$, an acceleration in lattice units), Hawking temperature $T_H=\kappa/(2\pi c_s)$, an
ergosphere at $r_e=\sqrt{A^2+B^2}/c_s$, and a superradiant threshold
$\omega=m\,\Omega_H$, $\Omega_H=Bc_s^2/A^2$. These values are the \emph{reference} against which the simulation output is read. The draining bathtub serves as a guide and a source of intuition: the object we construct is a lattice acoustic black hole with its own microphysics, whose kinematics we expect to be close, but not identical, to the continuum model.

The emergent horizon is located in two different ways (see Appendix~\ref{app:sec:horizon}). First by measuring where the azimuthally averaged radial
Mach profile $|\langle v_r\rangle_\varphi|/c_s$ crosses unity; for a rotating hole this differs from the surface $|\mathbf u|=c_s$, which is the ergosurface and lies outside it.
The steady Mach map [Fig.~\ref{fig:diag}(a)] develops a closed supersonic region, an annulus around the central absorber, whose outer boundary is an emergent $\mathrm{Ma}=1$ contour.
The azimuthally averaged profiles [Fig.~\ref{fig:diag}(b)] follow the known form
$|v_r|=A_{\rm eff}/r$ at the realised drain strength ($A_{\rm eff}/A\approx0.92$),
with $\rho\approx\mathrm{const}$, down to the near-horizon region, and the radial
mass flux $\Phi_M(r)=2\pi r\langle\rho v_r\rangle_\varphi$
[Fig.~\ref{fig:diag}(c)] is constant in $r$ near $-2\pi A_{\rm eff}$, confirming
that the emergent flow is a genuine steady draining solution. The measured horizon lies inside the reference radius $A_{\rm eff}/c_s$: between the two radii the flow sits on a subsonic plateau, and the crossing to $\mathrm{Ma}>1$ occurs at the transonic layer bounding the absorber, two to three cells thick; the horizon is anchored to the dissipation edge, and moving the absorber inward moves the horizon with it (Appendix~\ref{app:sec:horizon}). With rotation the horizon moves slightly inward and the ergosurface outward; the thin ergoregion between them is resolved, and the swirl $r\,v_\theta$ is constant outside the layer. At the resolution of Fig.~\ref{fig:diag} the sonic surface is circular.
At coarser resolution, four small supersonic islands appear in the transonic
layer, pinned to the lattice axes: they are the imprint of the $\mathrm{C}_4$
anisotropy of the square lattice, i.e., of its built-in breaking of rotation
(and hence Lorentz) invariance. They are causally inert and disappear as the
lattice is refined (a coarse-versus-fine comparison and a quantitative
anisotropy measure are shown in Appendix~\ref{app:sec:horizon}).

The complementary test, overlaid in Fig.~\ref{fig:diag}(a), is phonon ray tracing.
Acoustic null geodesics of the \emph{measured} metric are integrated
by fourth-order Runge--Kutta from Hamilton's equations for the
eikonal $H=c_s|\mathbf k|+\mathbf v\!\cdot\!\mathbf k$~\cite{Visser1998,BarceloLR,TorresJFM2018}:
\begin{equation}
\dot{\mathbf x}=c_s\hat{\mathbf k}+\mathbf v,\qquad
\dot k_a=-k_b\,\partial_a v_b,
\label{eq:rays}
\end{equation}
with $\mathbf v(\mathbf x)$
interpolated from the lattice. Rays launched inside the inner circle are trapped, those outside the outer circle escape. {Rays launched outwards from inside the outer circle, including from inside the annulus, are swept back and pile up at the inner circle}. The crossover radius coincides with the $\mathrm{Ma}=1$ surface. At $B=0$ and for radially launched rays the coincidence is exact, since then $\dot r=c_s+v_r$ vanishes where $\mathrm{Ma}=1$: the ray tracing checks the interpolated two-dimensional metric and shows the fate of the rays on either side of the sonic surface.

The surface gravity is where the lattice hole differs most from its continuum counterpart. We measure it two ways, both in lattice units (see Appendix~\ref{app:sec:horizon}). The first uses the exterior flow: fitting $|v_r|=A_{\rm loc}/r$ outside the transonic layer and extrapolating to the measured horizon gives $\kappa_{\rm flow}=c_sA_{\rm loc}/r_h^2=0.018\pm0.002$ at $A=17$ and $0.0215\pm0.0015$ at $A=11$. Both agree with the continuum reference $c_s^2/r_h$, which is $0.016$ and $0.023$ at these two sizes. The second uses the local gradient across the layer, $\kappa_{\rm layer}=|v_r\,dv_r/dr|_{r_h}$, which gives $0.11\pm0.02$ and $0.04\pm0.01$ for the same two sizes. The two measures differ because the sonic crossing is not sharp. It sits inside a transonic layer at the outer edge of the absorber, across which the density drops into the sink. That layer stays two to three cells thick at both hole sizes, so the gradient across it does not change with the hole while $\kappa_{\rm flow}$ falls as $1/A$: the ratio $\kappa_{\rm layer}/\kappa_{\rm flow}$ grows with the hole, from $\approx2$ at $A=11$ to $\approx6$ at $A=17$, the precise values depending on where within the layer the sonic crossing falls.
The exterior geometry and the horizon's own structure each carry a surface gravity. A Hawking temperature $T_H=\kappa/(2\pi c_s)$ is correspondingly not well defined (at least for this particular model) for this object: the two candidates differ by an order of magnitude at $A=17$, and which one a quantised field would see depends on the dispersion at the layer scale (Appendix~\ref{app:sec:horizon}), beyond the present classical study.

\begin{figure*}[t]
\centering
\includegraphics[width=\textwidth]{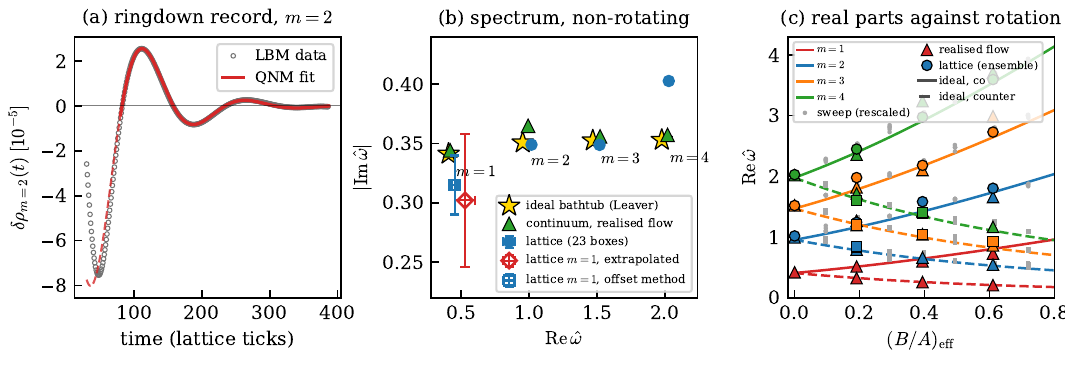}
\caption{\label{fig:qnm}
(a) A measured $m{=}2$ density ringdown on a probe ring (points) and its damped-sinusoid fit (line); the record is from one of the $23$ ensemble boxes of Table~\ref{app:tab:qnm}, on the probe ring at $1.4\,r_h$ where the kick is centred; the kick's own pulse occupies the first $80$ ticks, the fit is to the ringing that follows (solid) and is extrapolated back over the pulse (dashed), giving $\hat\omega=1.03-0.37\,i$.
(b) Non-rotating QNM spectrum in the complex plane: measured frequencies (circles, $23$-box mean and standard error) versus the continuum on the realised flow (triangles) and Leaver's continued fraction (stars).
{The two open symbols at $m{=}1$ are \emph{not} measurements but estimates, those of Table~\ref{app:tab:qnm}: extrapolated in $m$ (diamond) and offset from the realised flow (square).} (c) Rotating
draining bathtub: co-rotating (solid, circles) and counter-rotating (dashed,
squares) $\mathrm{Re}\,\hat\omega$ versus $(B/A)_{\rm eff}$ for $m=1$--$4$; lines: the ideal bathtub (Cardoso--Lemos--Yoshida~\cite{CLY2004}); large symbols: the ensembles of Tables~\ref{app:tab:qnm} and \ref{app:tab:rot} at $(B/A)_{\rm eff}=0$, $0.19$, $0.39$ and $0.61$, lattice (circles, squares) and continuum on the realised flow (triangles); small grey points: a sweep over $11$ rotations, five boxes each, with an earlier extractor and horizon convention, rescaled to the present horizon (Appendix~\ref{app:sec:qnmx}); it shows the trend only.}
\end{figure*}

\paragraph{Quasinormal frequencies.} Perturbed and released, a black hole rings down through a discrete set of
damped quasinormal modes with frequencies $\hat\omega=\hat\omega_R-i|\hat\omega_I|$ (in units of $c_s/r_h$),
fixed by the geometry alone. Here we can view this ringing by perturbing the fluid's steady density with a
localized Gaussian, record the response at probe rings, and project onto
azimuthal harmonics $e^{im\theta}$. The signal is a clean damped sinusoid
[Fig.~\ref{fig:qnm}(a)]. Fitting it returns a complex frequency, i.e., the quasinormal frequency.
We benchmark against the quasinormal frequencies of the draining-bathtub geometry computed exactly
by Leaver's continued-fraction method~\cite{Leaver1985}, and against the same wave equation on the realised flow, regular at the sonic point and outgoing at infinity (Appendix~\ref{app:sec:qnmx}).

Figure~\ref{fig:qnm}(b) shows the comparison against Leaver as computed in Refs.~\cite{CLY2004,BCL2004} and against the realised flow. Averaged over $23$ independent lattice boxes, the simulation gives the spectrum of Table~\ref{app:tab:qnm}, with the box-to-box standard error: the lattice and the realised flow ring higher than the ideal hole, whose horizon is larger; $m=2$ and $3$ damp alike in all three, the lattice $m=4$ faster. The offset is independent of the lattice viscosity (Appendix~\ref{app:sec:qnmx}).

Switching on rotation ($B>0$), the degenerate
modes split into co- and counter-rotating branches; Fig.~\ref{fig:qnm}(c) shows
the measured $\mathrm{Re}\,\hat\omega$ for $m=2,3,4$ tracking the
Cardoso--Lemos--Yoshida curves~\cite{CLY2004} across the full rotation range, and Table~\ref{app:tab:rot} gives the three spectra at three rotations, $(B/A)_{\rm eff}=0.19$, $0.39$ and $0.61$ as measured in the exterior flow, five boxes each, where the frequencies agree to about $10\%$ on both branches, with every mode damped, showing that the lattice analogue black hole, too, is classically
stable. The damping rates separate the realised flow from the ideal one: on the co-rotating branch the lattice and the realised flow damp faster than the bathtub, generally increasingly with rotation. On the counter-rotating branch the lattice modes damp more slowly than either continuum model at the two slower rotations.

The residual few-percent offsets are an effect of the near-horizon microphysics and of the finite lattice spacing, which acts as a short-distance cutoff; any \emph{physical} analogue differs from the ideal hole in this way, and the lattice hole is no exception.

The $m=1$ mode cannot be measured on the lattice: with $Q=\mathrm{Re}\,\hat\omega/2|\mathrm{Im}\,\hat\omega|\simeq0.60$ the $m=1$ mode never dominates the signal in any time window, so a damped-sinusoid fit is degenerate at any resolution. Both continuum spectra are computed at $m=1$; for the lattice we give two estimates (Table~\ref{app:tab:qnm}).

\begin{figure}[t]
\centering
\includegraphics[width=0.98\columnwidth]{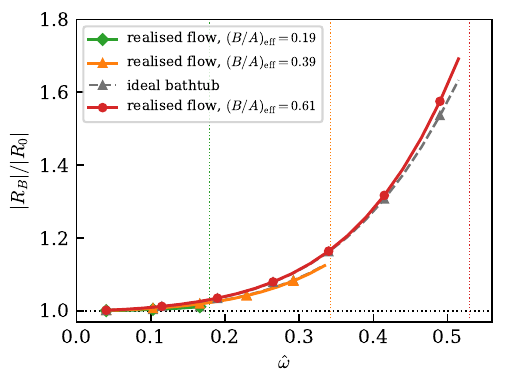}
\caption{\label{fig:sr}Superradiance of the realised flow: amplification excess $|R_B|/|R_0|$ against $\hat\omega$ ($m=1$), from loss-free scattering on the measured flows with the horizon absorbing in its own frame, at three rotations, up to the band edges $\Omega_H$ (dotted); dashed grey: ideal bathtub at the fitted circulation ($B/A=0.18$, $0.37$ and $0.56$, Appendix~\ref{app:sec:srad}). Reflection coefficients in Fig.~\ref{app:fig:srreflection}.}
\end{figure}

\paragraph{Superradiance and the horizon microphysics.} A rotating black hole can amplify waves: a mode $(\omega,m)$ scattered off the ergoregion returns with $|R|^2>1$ for $0<\omega<m\Omega_H$, extracting rotational energy~\cite{CLY2004,BasakMajumdar2003}. Whether the emergent geometry of the present model can support superradiance can be answered directly by reconstructing the metric from the measure of $\rho(r)$, $v_r(r)$ and $v_\theta(r)$ in the steady state, solve the loss-free radial scattering problem on the realised flow with purely ingoing conditions at the horizon, i.e. with the sonic point of the measured flow absorbing in its own frame, the horizon model of the previous section, and compare with the same calculation on the ideal bathtub. Without rotation the realised metric absorbs, $|R_0|^2<1$; with rotation it amplifies, $|R_B|^2>1$, in the band $0<\omega<m\Omega_H$, which closes at the measured $m\Omega_H$ to $1\%$: the peak gain $|R|^2-1$ reaches $6\%$ for $m=1$ at the fastest rotation (Fig.~\ref{app:fig:sr}), and the amplification excess $|R_B|/|R_0|$ tracks the Cardoso--Lemos--Yoshida result at the realised rotation to $4\%$ across the band [Fig.~\ref{fig:sr}]. In this sense the emergent geometry is a rotating acoustic black hole with an ergoregion; a reflective horizon fitted to the ringdown also amplifies (Appendix~\ref{app:sec:srad}).

Whether amplification appears dynamically is a different and the physically relevant question, dictated by the horizon microphysics. Superradiance is a statement about the rest frame of the absorber~\cite{Zeldovich1971,Zeldovich1972}: in that frame the wave has frequency $\omega'=\omega-m\Omega_{\rm abs}$ and is amplified when $\omega'<0$, because it then carries negative energy there and its absorption feeds the reflected wave; the same amplification by a rotating absorber has been measured acoustically~\cite{Cromb2020}. An ideal horizon absorbs in its own rotating frame, $\Omega_{\rm abs}=\Omega_H$. In the lattice hole the horizon is a layer supported by a dissipative sink fixed to the lattice, $\Omega_{\rm abs}=0$, so $\omega'=\omega>0$ for every mode and, on this argument, nothing is amplified. In linear driven runs the energy flux on rings next to the horizon is not conserved: it changes by large factors from ring to ring, and its sign does not follow the superradiant band and therefore cannot tell whether the hole amplifies. This seems to leave the question regarding superradiance unanswered. 

How amplification would appear as the horizon microphysics is varied can be understood within an effective description of the same scattering problem. We assign the horizon a single parameter (see Appendix~\ref{app:sec:srad}), a frame parameter $\alpha$, defined by letting the horizon absorb perfectly in a frame rotating at $\Omega_{\rm abs}=(1-\alpha)\Omega_H$, so that $\alpha=0$ is the ideal horizon and $\alpha=1$ the lattice sink. It enters the continuum scattering problem as a boundary condition near the horizon, and the flux balance then fixes the amplified band exactly: $|R_\infty|^2>1$ for $0<\omega<m(1-\alpha)\Omega_H$. The frame parameter narrows the band and reduces the gain together (for $m=1$ at $B/A=0.5$ on the ideal bathtub, from $5\%$ at $\alpha=0$ to $0.1\%$ at $\alpha=0.75$), until at $\alpha=1$ both senses of rotation are absorbed at every frequency. The quasinormal frequencies are nearly blind to $\alpha$: the co- and counter-rotating real parts shift by less than $7\%$ over $0\le\alpha\le0.75$, within the agreement of Fig.~\ref{fig:qnm}(c) with Cardoso--Lemos--Yoshida.
The ringdown is thus blind to the frame, while the position of the superradiant band edge, i.e. the frequency at which amplification switches off, measures $\alpha$. For an absorber placed several wavelengths inside the horizon, behind a supersonic buffer, or co-rotating with it, $\alpha\to0$ and the full gain of Fig.~\ref{fig:sr} is recovered.

The superradiance data do not constrain $\alpha$: the measured excess reflection $|R_B|/|R_0|$ for a co-rotating $m=1$ drive is one to within $\pm0.02$, in and above the band, but it was taken in the nonlinear regime.

\paragraph{Discussion.} We have shown that the kinematics of an analogue black hole emerges quantitatively from the local discrete rules of a lattice, and we have built one such example using the LB (D2Q9) scheme. The simulation reproduces the Cardoso--Lemos--Yoshida QNM spectra to a few percent without rotation and to about ten percent with it, with the residual offset traceable to the cell-scale structure of the horizon, i.e.\ a signature of the substrate lattice; the damping rates of the rotating hole are reproduced only in part by the two continuum models.
Superradiance separates into two ingredients that the lattice keeps apart. The emergent geometry amplifies in a loss-free calculation; a direct measurement on the running lattice does not show it and leaves the dynamical test inconclusive.
The frame in which the dissipation acts is, on a lattice, a design choice, which it is not in a water tank, a condensate or a photon fluid. In the model the lattice horizon is the sonic point of its own flow; whether it absorbs in the frame of the flow or in that of the lattice is what the superradiant band edge would measure.

The water-vortex experiments of Refs.~\cite{Torres2017,Torres2020}, where superradiant amplification was first measured, realise the same class of metrics through a depth-dependent wave speed and a nearly drain-free vortex; the lattice model we discussed here is closer to a compressible gas with a fixed sound speed and a dissipative drain. The two share the geometry, but seemingly not the horizon microphysics. 
Whether much more accurate (and computationally more intensive) numerical simulations could measure superradiance or whether dynamical superradiance on the lattice requires a more sophisticated model remains to be seen. 

The lattice substrate is interesting in its own right. It imprints a preferred frame: at coarse resolution the transonic
layer breaks into four supersonic islands locked to the
lattice axes, and the realised horizon carries two surface gravities, that of the exterior flow and that of the cell-scale transonic layer. The ringdown sees the same layer as an absorber whose damping depends on the mode and on its sense of rotation.
Were the substrate fundamental, horizon-scale structure of
this kind is what would distinguish it from a continuum
spacetime. Searching for such substrate-specific phenomena is a natural next use of the platform.


\section*{Acknowledgments} 
This work was conducted using the Miyabi Supercomputing System at the Joint Center for Advanced High Performance Computing (JCAHPC).
\noindent
\paragraph{Data availability and AI disclosure.} The code and data generated in this work are available at the Zenodo repository: https://doi.org/10.5281/zenodo.23660250. 
A Mathematica notebook that implements the model at lower accuracy is also included in the repository.\\ 
AI tools were used to aid with coding (sweeps, debugging, data analysis), graphics generation, and text polishing.

\appendix
\setcounter{secnumdepth}{2}

\section{The lattice-Boltzmann model}\label{app:sec:lbm}

The model we adopt is the standard D2Q9 lattice-Boltzmann scheme. Our implementation follows closely Refs.~\cite{Lallemand2000,Guo2002}. The method is not new and we report it only for completeness; the reader is addressed to the original references. The state of the system is described by nine real populations $f_i(\mathbf x,t)$, $i=0,\dots,8$, on a square lattice, advected along the discrete velocities $\mathbf e_i\in \{(0,0),(\pm1,0),(0,\pm1),(\pm1,\pm1)\}$ with weights $w_0=4/9$, $w_{1\text{--}4}=1/9$, $w_{5\text{--}8}=1/36$ and lattice sound speed $c_s=1/\sqrt3$. The macroscopic fields are
$\rho=\sum_i f_i$ and $\rho\mathbf u=\sum_i \mathbf e_i f_i$, and the second-order equilibrium is
\begin{equation}
f_i^{\rm eq}=w_i\rho\Big[1+3(\mathbf e_i\!\cdot\!\mathbf u)
+\tfrac92(\mathbf e_i\!\cdot\!\mathbf u)^2-\tfrac32|\mathbf u|^2\Big].
\end{equation}
The update is the collide rule
\begin{equation}
f_i^\star=f_i-[\mathsf M^{-1}\mathsf S\,\mathsf M(f-f^{\rm eq})]_i,
\end{equation}
followed by the stream rule
\begin{equation}
f_i(\mathbf x+\mathbf e_i,t{+}1)=f_i^\star(\mathbf x,t).
\end{equation}
The collision is of multi-relaxation-time type: $\mathsf M$ is the orthogonal moment matrix mapping
$f\mapsto(\rho,e,\varepsilon,j_x,q_x,j_y,q_y,p_{xx},p_{xy})$ and
$\mathsf S=\mathrm{diag}(s_a)$ holds the relaxation rates. With the velocity ordering $\mathbf e_0=(0,0)$;
$\mathbf e_{1,\dots,4}=(1,0),(0,1),(-1,0),(0,-1)$;
$\mathbf e_{5,\dots,8}=(1,1),(-1,1),(-1,-1),(1,-1)$, the moment matrix is the
standard integer matrix of Ref.~\cite{Lallemand2000}, rows ordered as the moments above,
\begingroup
\setlength{\arraycolsep}{3.6pt}
\begin{equation}
\mathsf M=\begin{pmatrix}
 1 & 1 & 1 & 1 & 1 & 1 & 1 & 1 & 1\\
-4 & -1 & -1 & -1 & -1 & 2 & 2 & 2 & 2\\
 4 & -2 & -2 & -2 & -2 & 1 & 1 & 1 & 1\\
 0 & 1 & 0 & -1 & 0 & 1 & -1 & -1 & 1\\
 0 & -2 & 0 & 2 & 0 & 1 & -1 & -1 & 1\\
 0 & 0 & 1 & 0 & -1 & 1 & 1 & -1 & -1\\
 0 & 0 & -2 & 0 & 2 & 1 & 1 & -1 & -1\\
 0 & 1 & -1 & 1 & -1 & 0 & 0 & 0 & 0\\
 0 & 0 & 0 & 0 & 0 & 1 & -1 & 1 & -1
\end{pmatrix},
\end{equation}
\endgroup
whose rows are mutually orthogonal. The relaxation matrix is
\begin{equation}
\mathsf S=\mathrm{diag}\big(0,\;s_e,\;s_\varepsilon,\;0,\;s_q,\;0,\;s_q,\;
s_\nu,\;s_\nu\big),
\end{equation}
with $s=0$ on the conserved moments $\rho,j_x,j_y$. The shear rate sets the viscosity $\nu=c_s^2(1/s_\nu-\tfrac12)$, with
$s_\nu=1/\tau$ and $\tau=0.70$ throughout. The ``ghost'' rate $s_q$ is fixed by the condition~\cite{Ginzburg2008}
\begin{equation}
\Big(\tfrac1{s_\nu}-\tfrac12\Big)\Big(\tfrac1{s_q}-\tfrac12\Big)
=\tfrac14,
\end{equation}
which sets $s_q=4/7\approx0.5714$ and, in our runs, eliminates the spurious near-boundary acoustic reflections that would otherwise contaminate the ringdown; the bulk rates are $s_e=1.1$, $s_\varepsilon=1.0$.

{To implement the body-force-core construction of the main text, the radial force $\mathbf F$, $F_r=g_r=-(A^2+B^2)/r^3$, is fixed in space and time and enters the collide rule through the second-order forcing scheme of Ref.~\cite{Guo2002}: the equilibrium moments are evaluated at the half-step velocity $\rho\mathbf u=\sum_i\mathbf e_i f_i+\tfrac12\mathbf F$ and the source $(\mathsf I-\tfrac12\mathsf S)\,\mathsf M F$ is added after relaxation, with $F_i=w_i\big[3\,\mathbf e_i\!\cdot\!\mathbf F+9(\mathbf e_i\!\cdot\!\mathbf u)(\mathbf e_i\!\cdot\!\mathbf F)-3\,\mathbf u\!\cdot\!\mathbf F\big]$. As described in Ref.~\cite{Guo2002} the prefactor $(\mathsf I-\tfrac12\mathsf S)$ removes the discrete-lattice error term of the Chapman--Enskog expansion, so that the recovered continuum equations are the isothermal Navier--Stokes equations with the force $\mathbf F$ added to the momentum balance and the viscosity unchanged; since $\sum_iF_i=0$ the force injects no mass. The $1/r^3$ profile is capped inside the absorber and tapered smoothly to zero across it and near the outer rim, so the force is exact throughout the region $r>r_h$ where the ringdown barrier lives. Being time independent, the force does not act on the perturbations, and the measured ringdown damping is not affected.}

\begin{figure*}[t]
\centering
\includegraphics[width=\textwidth]{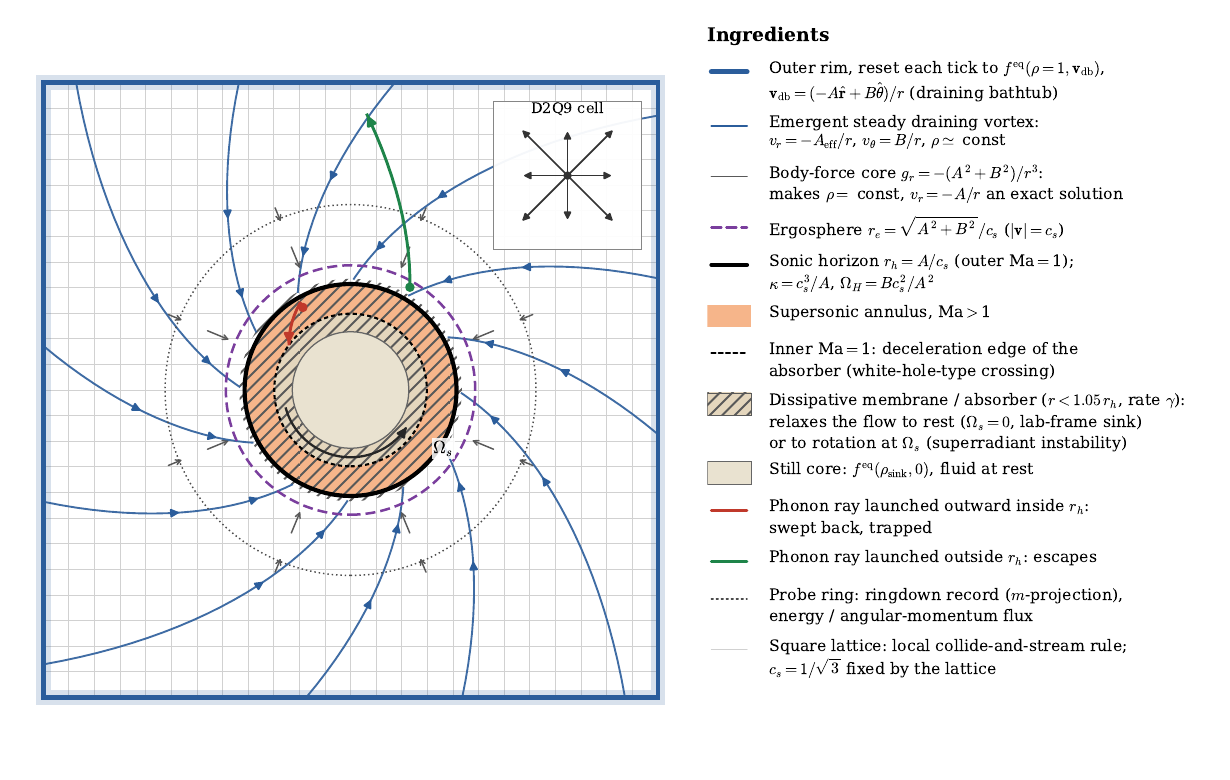}
\caption{\label{app:fig:schematic}
Schematic of the lattice black hole and of its ingredients (not to scale). From the outside in: the square D2Q9 lattice (grey grid; the collide-and-stream stencil is shown in the inset) with the outer rim reset at every step to the draining-bathtub state $f^{\rm eq}(\rho{=}1,\mathbf v_{\rm db})$; the emergent steady draining vortex (blue streamlines), sustained in the bulk by the body-force core $g_r=-(A^2+B^2)/r^3$ (grey arrows); the ergosphere $r_e$ (purple dashed), the sonic horizon $r_h=A/c_s$ (thick black, outer $\mathrm{Ma}=1$ crossing) and the supersonic annulus (orange) it bounds; the inner $\mathrm{Ma}=1$ crossing (black dashed), i.e., the deceleration edge of the absorber; the dissipative membrane (hatched, $r\lesssim r_h$), which relaxes the flow to rest (the laboratory-frame sink); and the still core, relaxed to $f^{\rm eq}(\rho_{\rm sink},0)$, where the fluid does not move. Also shown are a phonon ray launched outward inside $r_h$ and swept back (red), one launched outside $r_h$ that escapes (green), and one of the probe rings (dotted) on which the ringdown and the energy fluxes of the driven waves are recorded.}
\end{figure*}

The central absorber relaxes the populations at every step towards $f^{\rm eq}(\rho_{\rm sink},0)$, $\rho_{\rm sink}=0.05$, with a weight equal to one for $r\le0.6\,A/c_s$ and falling linearly to zero over the next three cells, so that the measured horizon lies at the outer end of this ramp; the rim reset acts for $r>R=0.45\,N$ on an $N^2$ lattice, and each run is settled for $2400\,(R/110)^2$ steps and monitored for as many again. The full construction is summarised in Fig.~\ref{app:fig:schematic}.

\section{Ray tracing and horizon structure}\label{app:sec:horizon}

\emph{Steady state.} After relaxation the fields are followed for twice as long: the largest density change per step falls to round-off, and the ring-averaged $\rho$, $v_r$, $v_\theta$ are constant to five digits, with and without rotation. The steady state is the same in every box, so the horizon radius and the exterior profiles do not depend on the domain size.

\emph{Location.} The horizon is the outermost radius where the angle-averaged radial velocity, sampled on rings at sub-cell spacing, reaches the sound speed; the ergosurface is where $|\mathbf u|=c_s$. The swirl $r\,v_\theta$ is constant outside the transonic layer, and $\Omega_H=v_\theta/r$ is read at the horizon. The band edges so measured, $\hat\Omega_H=\Omega_Hr_h/c_s=0.180$, $0.343$ and $0.529$ at $(B/A)_{\rm eff}=0.19$, $0.39$ and $0.61$, fall below $(B/A)_{\rm eff}$ because the swirl drops by $15$--$25\%$ across the transonic layer. The figures use one-cell shells, which agree with the ring sampling to a fraction of a cell.

\emph{Ray tracing.} Since $H$ is homogeneous of degree one in $\mathbf k$, the path depends only on the direction $\hat{\mathbf k}$ and the initial $|\mathbf k|$ is arbitrary.
The integration is fourth-order Runge--Kutta with step $0.4$ lattice ticks, up to $4000$ steps per ray.

Rays are launched just outside the absorber with $\mathbf k$ pointing radially \emph{outward}, and terminated when they enter the drain region (absorbed) or reach the box rim (escaped).

Rays launched \emph{inside the supersonic annulus} with $\mathbf k$ pointing outward are carried inward and stall at the \emph{inner} $\mathrm{Ma}=1$ crossing; a ray launched outward from the core crosses the still fluid at $c_s$ and stalls on the same surface from within. The inner crossing is a white-hole--type horizon for phonons (outgoing characteristics \emph{converge} on it, wavefronts pile up and blueshift), while the outer crossing is the black-hole horizon.
The pile-up at the inner crossing is regularized by the absorber, which erases the rays.

\begin{figure*}[t]
\centering
\includegraphics[width=0.78\textwidth]{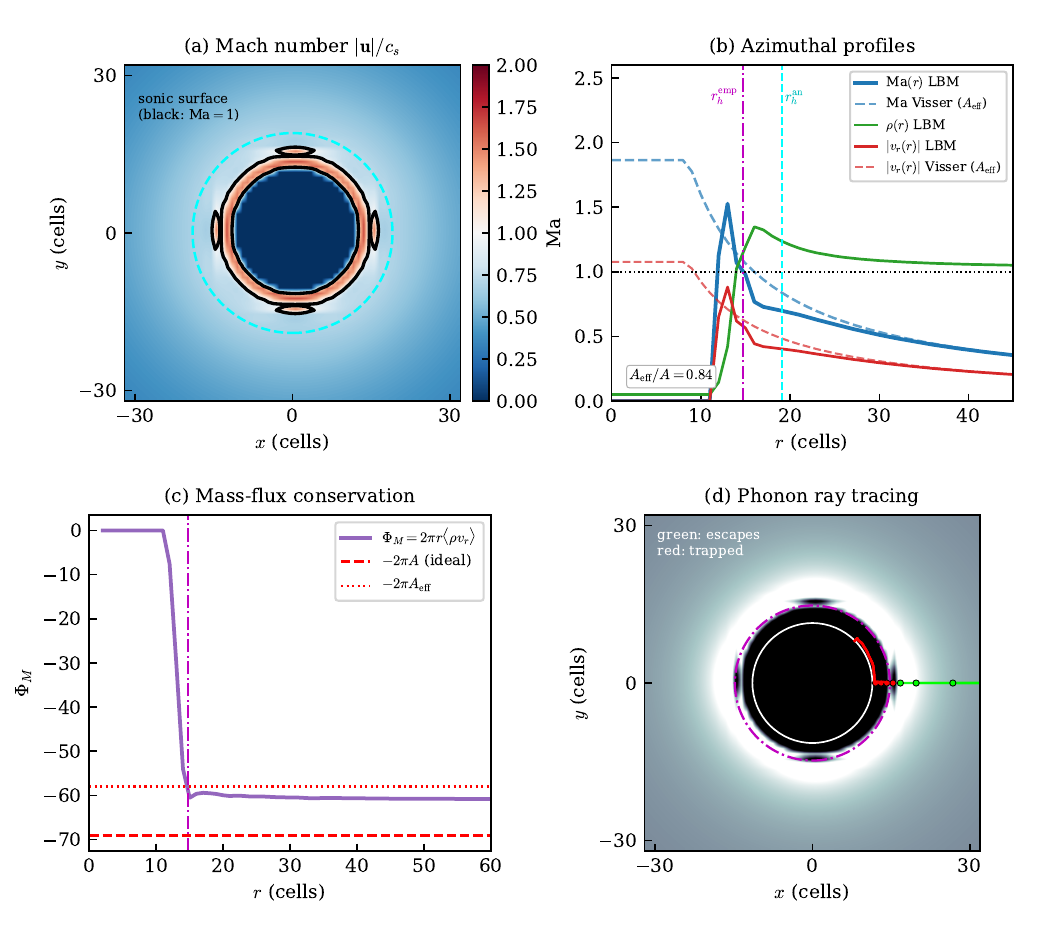}
\caption{\label{app:fig:fig1coarse}
Diagnostics of Fig.~\ref{fig:diag} rederived on the coarse run
($A=11$, $B=0$, $300^2$ cells, body-force core; $r_h^{\rm emp}=14.8$). (a) Mach
map (radial Mach number at $B=0$). The four small axis-pinned
islands are the $\mathrm{C}_4$ lattice anisotropy of the transonic
layer, averaged over in (b) and causally inert in (d); the dashed cyan circle
is the ideal $r_h=A/c_s$. (b) Azimuthal profiles versus
the Visser form \cite{Visser1998} at the realised $A_{\rm eff}$ ($A_{\rm eff}/A\approx0.84$).
(c) Mass-flux plateau at $-2\pi A_{\rm eff}$. (d) Phonon ray tracing on the
measured metric: launch radii $12.0$--$15.1$ trapped (red),
$16.5$--$39.5$ escape (green).}
\end{figure*}

\emph{Surface gravity.} The two surface gravities quoted in the main text are measured as follows.
For $\kappa_{\rm flow}$ we fit $|v_r|=A_{\rm loc}/r$ over windows running
from $r_h+2$ to $r_h+20$ cells and evaluate the Visser form \cite{Visser1998} at the measured
$r_h$; the quoted spread is the window dependence. For $\kappa_{\rm layer}$
we take the local gradient of the azimuthally averaged profile across the
sonic crossing with one- and two-cell differencing windows. The coarse value
is additionally affected by the axis islands of
Fig.~\ref{app:fig:fig1coarse}.

The profiles show where the difference comes from. Outside the horizon the flow follows $|v_r|=A_{\rm eff}/r$ at constant density, with $A_{\rm eff}$ below $A$ because the accretion rate is set at the outer reservoir ($A_{\rm eff}/A=0.84$ in the coarse run and $0.92$ in the fine one). Outside the sonic surface
the fine run has a subsonic plateau, $\mathrm{Ma}\simeq0.82$--$0.84$ at
$\rho\simeq1.2$--$1.5$, reaching out to $r\simeq1.4\,r_h$; the crossing to
$\mathrm{Ma}>1$ then happens in a layer at the outer edge of the absorber,
across which the density falls from $\simeq1.4$ to $\simeq0.14$. Measured
from the innermost $\mathrm{Ma}=1.3$ crossing to the outermost $\mathrm{Ma}=0.95$ crossing on the one-cell shells of the figures, that layer is $2.3$ cells thick in the
fine run and $2.7$ in the coarse one. The same structure pulls the horizon
inside the reference radius $A_{\rm eff}/c_s$, and ties it to the absorber: with a second absorber deeper in the drain, at rest or co-moving, the horizon moved onto its edge in every stable configuration. The sink cannot signal upstream through the supersonic flow, so the fluid goes transonic wherever removal begins; horizon and dissipation edge are one object, and the ringdown (Sec.~\ref{app:sec:qnmx}) is decided in that layer.

Refining the lattice does not bring the two values together. The layer is a
fixed number of cells, set by the competition between the sink relaxation
and the streaming step, while $\kappa_{\rm flow}\propto c_s^3/A_{\rm eff}$
falls as the hole grows in cells. The ratio grows without bound,
and the continuum limit of this object is a horizon that is a step on the
scale of its own exterior geometry.

Which of the two a field feels is then a selection
by wavelength. A wave samples the flow over its own $\lambda$: modes with
$\lambda$ well above the layer width see the exterior gradient, modes with
$\lambda$ comparable to it see the step. For Hawking radiation the relevant
wavelength is the dispersion scale of the
substrate~\cite{Unruh1995,BroutMassar1995,CorleyJacobson1996,UnruhSchutzhold2005},
which on this lattice is the cell — the same scale as the layer. A quantised
field here would then be expected to see something close to $\kappa_{\rm layer}/(2\pi c_s)$,
unless the dispersion scale were pushed well above the layer width. This is
the usual dispersive result: the temperature follows the gradient over the
region where dispersion acts, and reduces to the surface gravity of the
exterior geometry only when that region is small compared with the scale on
which the flow varies. The lattice horizon violates that condition by
construction.

So $\kappa_{\rm layer}$ is a property of the substrate,
whereas $r_h$, the exterior flow and the quasinormal spectrum are set at
scales $\gtrsim r_h$ and track the continuum reference.

\begin{figure*}[t]
\centering
\includegraphics[width=0.78\textwidth]{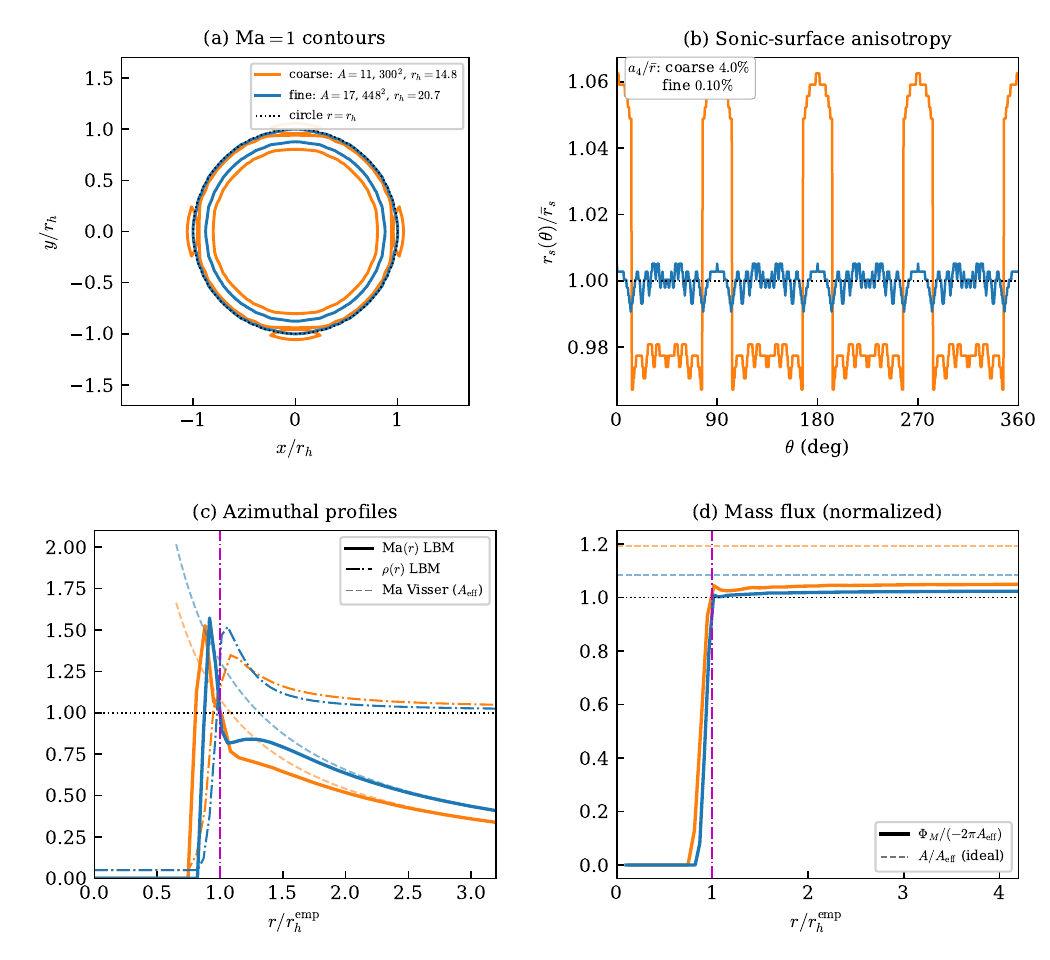}
\caption{\label{app:fig:fig1overlay}
Horizon-scaled overlay of the coarse (orange; $A=11$, $300^2$,
$r_h^{\rm emp}=14.8$; Fig.~\ref{app:fig:fig1coarse}) and fine
(blue; $A=17$, $448^2$, $r_h^{\rm emp}=20.7$; Fig.~1 of
the main text) runs. (a)
$\mathrm{Ma}=1$ contours in units of $r_h^{\rm emp}$: the surfaces coincide except for the four axis ovals and the waviness of the coarse contour.
(b) Sonic-surface radius versus angle: the coarse surface is a fourfold square wave, $a_4/\bar r_s=4.0\%$, where $\bar r_s$ is the angular mean and $a_4 = 2 \left|\langle r_s(\theta) e^{-4 i \theta} \rangle_\theta\right|$; the fine surface is flat to the pixel level ($a_4/\bar r_s=0.10\%$). (c) Azimuthal Mach and density profiles against $r/r_h^{\rm emp}$, with the Visser $A_{\rm eff}/r$ forms (thin dashed). (d) Mass flux normalized to $-2\pi A_{\rm eff}$: both plateaus
jump at $r/r_h^{\rm emp}=1$ and lie flat near unity; thin dashed lines mark
the nominal $A/A_{\rm eff}$.}
\end{figure*}

At coarse resolution the transonic layer exhibits four symmetric supersonic islands pinned to the lattice axes. Figure~\ref{app:fig:fig1coarse} is the analogue of Fig.~\ref{fig:diag} for the coarse run ($A=11$, $300^2$ cells), where the islands are visible in the Mach map. Figure~\ref{app:fig:fig1overlay} overlays the coarse and fine sonic surfaces and profiles in horizon units to quantify the anisotropy.

\section{Quasinormal-mode extraction}\label{app:sec:qnmx}

To measure the ringdown we perturb the steady density by a localized Gaussian, record the response on probe rings, project it onto $e^{im\theta}$, and obtain the complex frequencies by a non-linear damped-sinusoid fit, converted to horizon units $\hat\omega=\omega\,r_h^{\rm emp}/c_s$.

Table~\ref{app:tab:qnm} lists the non-rotating frequencies against the results of Ref.~\cite{BCL2004} and against the continuum acoustics on the realised flow.

\begin{table}[t]
\caption{\label{app:tab:qnm} Non-rotating quasinormal frequencies $\hat\omega=\omega r_h/c_s$, $r_h=14.40$: lattice (body-force core, $A=11$; mean over $23$ independent runs of different size, box-to-box standard error in parentheses; kick $2\times10^{-4}$), continuum acoustics on the realised flow, and the ideal bathtub (Leaver, Ref.~\cite{BCL2004}). The two $m=1$ lattice entries are the estimates described below: the extrapolation in $m$ (upper) and the offset from the realised flow (lower).}
\begin{ruledtabular}
\begin{tabular}{ccccccc}
$m$ & $\mathrm{Re}\,\hat\omega_{\rm LBM}$ & $\mathrm{Re}\,\hat\omega_{\rm flow}$ & $\mathrm{Re}\,\hat\omega_{\rm L}$ &
$|\mathrm{Im}\,\hat\omega_{\rm LBM}|$ & $|\mathrm{Im}\,\hat\omega_{\rm flow}|$ & $|\mathrm{Im}\,\hat\omega_{\rm L}|$\\
\colrule
1 & $0.53(7)$ & $0.419$ & $0.407$ & $0.30(6)$ & $0.344$ & $0.341$\\
 & $0.45(2)$ & & & $0.32(3)$ & & \\
2 & $1.017(4)$ & $0.992$ & $0.953$ & $0.349(1)$ & $0.365$ & $0.351$\\
3 & $1.519(12)$ & $1.522$ & $1.469$ & $0.349(3)$ & $0.356$ & $0.352$\\
4 & $2.028(3)$ & $2.019$ & $1.976$ & $0.403(2)$ & $0.357$ & $0.353$\\
\end{tabular}
\end{ruledtabular}
\end{table}

\emph{Continuum on the realised flow.} Linearised isothermal acoustics for a potential perturbation $\phi(r)e^{i(m\theta-\omega t)}$ on the measured background $\rho_0(r)$, $v_r(r)$, $v_\theta(r)$ gives $A\phi''+B\phi'+C\phi=0$ with $A=v_r^2-c_s^2$, $B=-2i\varpi v_r+[(rWv_r^2)'-(r\rho_0)']/(rW)$, $C=-\varpi^2-i(rWv_r\varpi)'/(rW)+m^2c_s^2/r^2$, $W=\rho_0/c_s^2$ and $\varpi=\omega-mv_\theta/r$. The measured profiles are used inside $2.9\,r_h$ and the fitted ideal flow outside; the solution regular at the sonic point (the $|v_r|=c_s$ crossing, as for the lattice horizon) is integrated outward, the outgoing one inward along a complex-$r$ ray, and the quasinormal frequency is the zero of their Wronskian.

\emph{Two-dimensional check and viscosity.} The same equations, solved in the time domain on the measured two-dimensional background of one box with the lattice's kick, absorber, rim and extraction, reproduce the one-dimensional spectrum without viscosity; the four-fold background adds only the $m=2$ cos/sin split of any square grid. With the lattice's nominal viscosities (shear $\nu=0.067$, bulk $0.136$) the Navier--Stokes twin damps $m=4$ by $0.47$, against $0.41$ on the lattice and $0.32$--$0.35$ without viscosity, mostly by attenuation in the exterior; the lattice damping does not move with the viscosity ($0.41$ at $\nu=0.033$ and $0.067$, $0.40$ with the bulk rate halved), and sweeping the lattice viscosity ($\tau$ between $0.56$ and $0.90$) leaves the spectrum unchanged, which excludes the bulk.

\emph{Effective reflective horizon.} The spectrum also has a compact description on the ideal bathtub, with the lattice horizon entering through one complex reflectivity, $H\to M_-+R_h\,M_+$, $M_\mp$ being the ingoing and outgoing horizon solutions. The $m=3$ and $4$ modes of Table~\ref{app:tab:qnm} give close values, $0.26\,e^{i132^\circ}$ and $0.33\,e^{i128^\circ}$; their mean, $R_h=0.30\,e^{i130^\circ}$, reproduces both to $1.3\%$ and $m=2$ to $3\%$ in the real part and $10\%$ in the damping, comparable to the realised flow. The two descriptions are complementary: $R_h$ condenses into one number how the realised hole differs from the ideal one, chiefly by its horizon lying inside the ideal radius, which the realised flow contains explicitly and without free parameters; on the realised flow the reflectivity is small, $|R_h|^2<0.3\%$.

While measuring $m>1$ is straightforward, measuring the $m=1$ mode was not possible at least for the parameter window and resolution considered. The $m=1$ mode is trapped by a centrifugal barrier five times shallower than at $m=2$, so its quality factor $Q=\mathrm{Re}\,\hat\omega/2|\mathrm{Im}\,\hat\omega|$ is only $Q\simeq0.6$ and the mode decays to $e^{-\pi/Q}\simeq5\times10^{-3}$ of its amplitude within one cycle. With less than one oscillation visible, a damped-sinusoid fit is degenerate.
This is a property of the geometry: the same shallowness renders the WKB expansion non-convergent at $m=1$~\cite{BCL2004}.

It is possible to extrapolate the value at $m=1$. Since Leaver's continued fraction does converge at $m=1$, giving $\hat\omega_{m=1}=0.406887-0.341307\,i$, an extrapolation can be calibrated. We fit the measured $m=2,3,4$ fundamentals in $m$ with two-parameter models ($am+b$, $am+c/m$, $am+c/m^2$ for the real part; $b$, $am+b$, $b+c/m$ for the damping) and evaluate at $m=1$; applying the same fit to Leaver's own $m=2,3,4$ and comparing with his exact $m=1$ measures the estimator's bias, which we subtract. The quoted uncertainty has two parts, added in quadrature. The first is statistical: we resample the $m=2,3,4$ frequencies within their ensemble spreads and repeat the fit, and take the spread of the resulting $m=1$ values. The second is systematic: the models are averaged, and the spread across models is carried as the systematic error. The fit is restricted to two free parameters: with three data points a three-parameter form interpolates the scatter exactly and then extrapolates it.
The result is $\hat\omega_{m=1}=0.528(74)-0.302(56)\,i$.

A second estimate adds to the realised-flow $m=1$ the lattice--realised-flow offset at $m=2$ and $3$ ($+0.03+0.02\,i$ and $0.00+0.01\,i$), extrapolated to $m=1$: $\hat\omega_{m=1}=0.45(2)-0.32(3)\,i$. Neither estimate is a measurement.

For the rotating case the modes split into co- ($+m$) and counter-rotating
($-m$) branches. The rotating ensemble has five boxes of $248$--$372$ cells per rotation, kicked at $2\times10^{-4}$ in $e^{\pm im\theta}$ channels; the horizon radii are those of Table~\ref{app:tab:rot} and $(B/A)_{\rm eff}=\langle r v_\theta\rangle/\langle -r v_r\rangle$ is read from the exterior flow over $1.2$--$2.8\,r_h$. Table~\ref{app:tab:rot} gives the three spectra: lattice, continuum on the realised rotating profiles (the solver of this section with the measured $v_\theta$), and the ideal bathtub, i.e. the
Cardoso--Lemos--Yoshida values~\cite{CLY2004} at the realised $(B/A)_{\rm eff}$.
 Left undisturbed, the same boxes show every $m=1$--$4$ channel decaying, so no instability contaminates the records.

\begin{table}[t]
\caption{\label{app:tab:rot}Rotating quasinormal frequencies $\hat\omega=\omega r_h/c_s$ at the three measured rotations: lattice (five boxes, box-to-box standard error in parentheses; --- where the extraction does not resolve the mode), continuum on the realised flow, and ideal bathtub at the same $(B/A)_{\rm eff}$. Positive $m$ co-rotating, negative counter-rotating.}
\begin{ruledtabular}
\begin{tabular}{rccc}
\multicolumn{4}{c}{$(B/A)_{\rm eff}=0.19$ ($r_h=14.35$)} \\
$m$ & lattice & realised flow & ideal \\
\colrule
$+1$ & --- & $0.514-0.391\,i$ & $0.510-0.370\,i$ \\
$-1$ & --- & $0.322-0.305\,i$ & $0.324-0.310\,i$ \\
$+2$ & $1.241(12)-0.459(5)\,i$ & $1.195-0.436\,i$ & $1.160-0.381\,i$ \\
$-2$ & $0.838(30)-0.253(8)\,i$ & $0.793-0.351\,i$ & $0.783-0.317\,i$ \\
$+3$ & $1.979(6)-0.412(26)\,i$ & $1.817-0.477\,i$ & $1.781-0.384\,i$ \\
$-3$ & $1.199(9)-0.230(7)\,i$ & $1.266-0.365\,i$ & $1.211-0.318\,i$ \\
$+4$ & $2.449(23)-0.348(25)\,i$ & $2.356-0.589\,i$ & $2.394-0.384\,i$ \\
$-4$ & $1.601(2)-0.261(2)\,i$ & $1.717-0.353\,i$ & $1.632-0.318\,i$ \\
\colrule
\multicolumn{4}{c}{$(B/A)_{\rm eff}=0.39$ ($r_h=14.29$)} \\
$m$ & lattice & realised flow & ideal \\
\colrule
$+1$ & --- & $0.599-0.468\,i$ & $0.642-0.397\,i$ \\
$-1$ & --- & $0.257-0.278\,i$ & $0.258-0.277\,i$ \\
$+2$ & $1.582(8)-0.479(10)\,i$ & $1.432-0.521\,i$ & $1.421-0.409\,i$ \\
$-2$ & --- & $0.661-0.301\,i$ & $0.641-0.281\,i$ \\
$+3$ & $2.181(8)-0.474(24)\,i$ & $2.104-0.625\,i$ & $2.171-0.411\,i$ \\
$-3$ & $1.041(35)-0.210(13)\,i$ & $1.044-0.300\,i$ & $0.995-0.282\,i$ \\
$+4$ & $2.977(37)-0.616(40)\,i$ & $3.239-0.602\,i$ & $2.913-0.412\,i$ \\
$-4$ & $1.400(9)-0.262(19)\,i$ & $1.413-0.287\,i$ & $1.342-0.282\,i$ \\
\colrule
\multicolumn{4}{c}{$(B/A)_{\rm eff}=0.61$ ($r_h=14.24$)} \\
$m$ & lattice & realised flow & ideal \\
\colrule
$+1$ & --- & $0.727-0.534\,i$ & $0.804-0.420\,i$ \\
$-1$ & --- & $0.206-0.246\,i$ & $0.207-0.244\,i$ \\
$+2$ & $1.803(15)-0.494(16)\,i$ & $1.659-0.598\,i$ & $1.737-0.431\,i$ \\
$-2$ & --- & $0.546-0.262\,i$ & $0.526-0.245\,i$ \\
$+3$ & $2.729(16)-0.672(38)\,i$ & $2.991-0.721\,i$ & $2.644-0.434\,i$ \\
$-3$ & $0.927(9)-0.251(10)\,i$ & $0.862-0.258\,i$ & $0.819-0.245\,i$ \\
$+4$ & $3.595(13)-0.887(25)\,i$ & $3.710-0.667\,i$ & $3.543-0.435\,i$ \\
$-4$ & --- & $1.166-0.246\,i$ & $1.105-0.245\,i$ \\
\end{tabular}
\end{ruledtabular}
\end{table}

\begin{figure*}[t]
\centering
\includegraphics[width=0.95\textwidth]{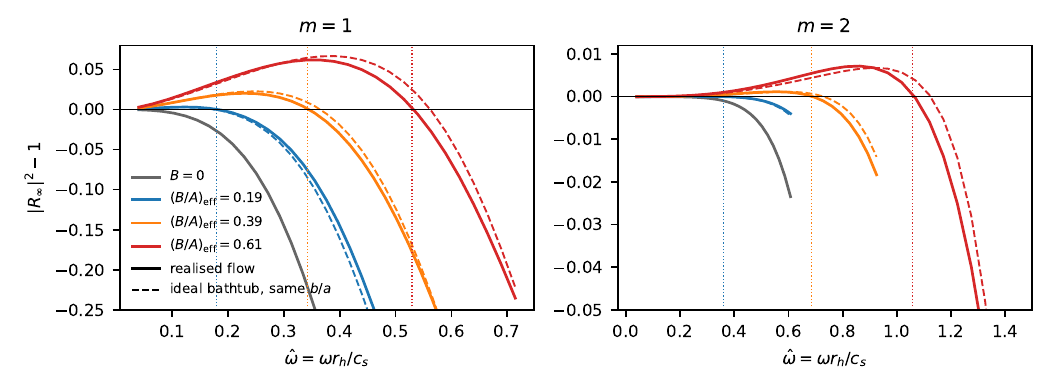}
\caption{\label{app:fig:sr}Superradiant gain $|R_\infty|^2-1$ of the realised flows (solid) against $\hat\omega$, $m=1$ (left) and $m=2$ (right), horizon absorbing in its own frame; dashed: ideal bathtub at the same fitted circulation ($B/A=0.18$, $0.37$ and $0.56$); dotted: the measured $m\Omega_H$. Without rotation the flow absorbs at every frequency.}
\end{figure*}

\begin{figure*}[t]
\centering
\includegraphics[width=0.92\textwidth]{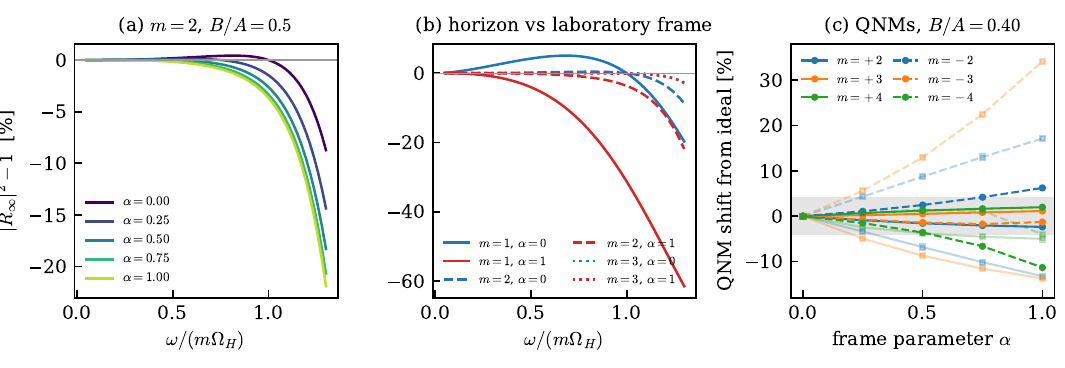}
\caption{\label{app:fig:law}
Superradiance under the absorption-frame law, Eq.~\eqref{app:eq:law}, on the draining bathtub. (a) Gain spectrum of $m=2$ at $B/A=0.5$ for $\alpha=0$--$1$. (b) Horizon-frame ($\alpha=0$) versus laboratory-frame ($\alpha=1$) absorber for $m=1,2,3$. (c) Shift of the co- (solid) and counter-rotating (dashed) quasinormal frequencies at $B/A=0.40$ from their ideal values as a function of $\alpha$ (circles: real part; faint squares: damping); the grey band marks $\pm4\%$. All panels are the law on the analytic bathtub with $\alpha$ as an external parameter.}
\end{figure*}

\section{Superradiance of the emergent geometry}\label{app:sec:srad}
\begin{figure}[t]
\centering
\includegraphics[width=0.95\columnwidth]{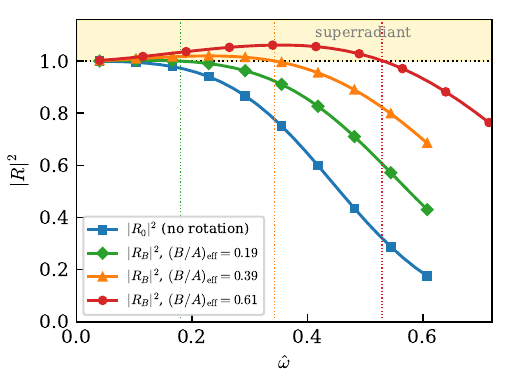}
\caption{\label{app:fig:srreflection}{%
Energy reflection $|R|^2$ versus $\hat\omega$ from conservative
scattering on the measured flows ($m=1$), with the horizon absorbing in its own frame: the non-rotating
channel absorbs ($|R_0|^2<1$), rotation lifts the reflection
above unity ($|R_B|^2>1$; shaded band, for $\omega<\Omega_H$, dotted verticals). The amplification excess
$|R_B|/|R_0|$ of these curves is compared with ideal-bathtub theory in
Fig.~\ref{fig:sr}.}}
\end{figure}

 We have extracted from the steady state $\rho(r)$, $v_r(r)$, $v_\theta(r)$, and solved the loss-free radial scattering problem{. Writing the
perturbation of the velocity potential in the mode form
$\psi(t,r,\theta)=r^{-1/2}\,H(r_\ast)\,e^{i(m\theta-\omega t)}$, the wave
equation on the emergent acoustic geometry separates into a radial
equation for the wavefunction $H$ (see Refs.~\cite{BCL2004,CLY2004}),}
$$
H''+[\,(\omega-m\Omega(r))^2-V(r)\,]H=0
$$
{where primes denote derivatives with respect to the tortoise
coordinate $r_\ast$, defined by $dr_\ast/dr=(1-v_r^2/c_s^2)^{-1}$ (it
sends the horizon to $r_\ast\to-\infty$), and
$V(r)=(1-v_r^2/c_s^2)\big[(m^2-\tfrac14)/r^2
+\tfrac54\,v_r^2/(c_s^2r^2)\big]$ is the curvature (centrifugal)
potential,} with $\Omega=v_\theta/r$ and purely ingoing boundary conditions at the horizon ($k_H=\omega-m\Omega_H${, with $\Omega_H=Bc_s^2/A^2$ the angular velocity of the hole}).
On the realised flow, whose density varies near the hole, the problem is solved as in Sec.~\ref{app:sec:qnmx}: ingoing at the sonic point, i.e.\ absorbing in its own frame, with the exterior continued into the fitted ideal flow. The ideal curves of Figs.~\ref{fig:sr} and \ref{app:fig:sr} are the bathtub at the circulation fitted to the exterior of each realised flow, $B/A=0.18$, $0.37$ and $0.56$, within $6\%$ of the measured band edges, whereas the ideal column of Table~\ref{app:tab:rot} is the bathtub at $(B/A)_{\rm eff}$ itself. For real $\omega$ the radial energy current $J=r\,(\rho_0/c_s^2)\,[\,(c_s^2-v_r^2)\,{\rm Im}(\phi^*\phi')+v_r\varpi|\phi|^2]$ is conserved, and at large $r$ the solution is decomposed on in- and outgoing WKB waves of unit current, giving $|R_\infty|^2=J_{\rm out}/|J_{\rm in}|=1+J_{\rm hor}/|J_{\rm in}|$. On the ideal bathtub the solver reproduces the reflection coefficients of Refs.~\cite{BCL2004,CLY2004} to four digits. On the realised profile the solver gives $|R_0|^2<1$ without rotation and $|R_B|^2>1$ with rotation {[Fig.~\ref{app:fig:srreflection}]}; Fig.~\ref{app:fig:sr} gives the gain $|R_\infty|^2-1$ for $m=1$ and $2$, and the amplification excess $|R_B|/|R_0|$ is Fig.~\ref{fig:sr}.

While this indicates that the geometry realised by the flow can support amplification, it leaves open the question of whether amplification shows up in the lattice model dynamically and can be observed directly by measuring the reflected amplitude. The observable is the excess $|R_B|/|R_0|$ of the rotating over the non-rotating reflection, from the same in/out decomposition as above, for a co-rotating $m=1$ drive. At $A=11$ [$(B/A)_{\rm eff}=0.605$, band edge at the measured $\hat\Omega_H\simeq0.53$] it is $1.0056$, $0.9985$, $1.0046$ at $\hat\omega=0.20$, $0.30$, $0.40$, where the conservative calculation on the same realised metric gives $1.039$, $1.117$, $1.283$; the scatter across $A=11$, $16$, $22$ is $\pm0.01$. At $A=17$ [$1064^2$, $(B/A)_{\rm eff}=0.63$] the in-band points are $1.006$, $0.998$, $1.005$ ($\pm0.007$), and $\hat\omega=0.50$, $0.65$, $0.70$, $0.80$, $0.90$ give $0.998$, $1.010$, $1.007$, $0.998$, $0.979$ against conservative values $1.60$ to $4.33$. In units of $|R|^2-1$ the floor of the measurement is $\pm0.02$. These values do not test linear scattering. The drive, applied on a ring at every step, builds a steady response $\delta\rho\simeq0.05$--$0.09$ with harmonics at $\epsilon=1.5\times10^{-3}$, and just outside the horizon the flux of the driven harmonic does not scale as $\epsilon^2$ between $\epsilon=0.012$ (the value used above) and $7.5\times10^{-4}$, at $m=1,2$ and $A=11,17$. The response is linear at $\epsilon=10^{-5}$. There the energy flux $I=\langle(p'+\rho_0\mathbf v_0\!\cdot\!\mathbf v')(\mathbf v'+\rho'\mathbf v_0/\rho_0)\rangle$ of the driven $m=1$ harmonic on rings $r_h+1$ to $r_h+4$ is outward for the co-rotating wave at every frequency up to twice the band edge, with no feature at the edge, and inward for non-rotating and counter-rotating waves; but it changes by factors between adjacent rings and grows $5$--$7$-fold out to $r=80$. Next to the gas horizon, where the flow varies on the cell scale, the quadratic flux is not conserved and its sign does not measure what the hole does: the dynamical test is not available, at either hole size. The continuum twin of the rotating flow, driven and read out the same way, does not conserve this flux either, while both conserve it without rotation: the deficiency is in the flux expression on the swirling flow, not in the lattice.

\subsection{Superradiance and the absorption frame}\label{app:sec:law}
Here we ask how superradiance would appear if the dissipation acted not in the horizon's own frame but in another, as a sink fixed to the lattice might. We work on the ideal draining bathtub with circulation, in units $r_h=A=c_s=1$ ($\Omega_H=B$), where the radial equation is $d^2H/dr_\ast^2+[(\omega-mB/r^2)^2-V]H=0$ with $V$ as above; near the horizon $H=a\,M_-+b\,M_+$, $M_\mp\propto x^{\mp i\varpi/2}$, $x=r-1$, $\varpi=\omega-m\Omega_H$, and the ideal horizon is $b=0$. A boundary that absorbs perfectly in a frame rotating at $\Omega_{\rm abs}=(1-\alpha)\Omega_H$ obeys the impedance condition $dH/dr_\ast=-i\varpi_{\rm eff}H$ with $\varpi_{\rm eff}=\omega-m(1-\alpha)\Omega_H$, i.e.
\begin{equation}
\frac{b}{a}=\frac{\varpi-\varpi_{\rm eff}}{\varpi+\varpi_{\rm eff}},
\label{app:eq:law}
\end{equation}
which is the ideal horizon for $\alpha=0$ and, for $\alpha=1$, the boundary absorbs in the laboratory frame, as the lattice sink does. Conservation of the Wronskian current gives the exact flux identity $\omega(|R_\infty|^2-1)=\varpi(|b|^2-|a|^2)/|I|^2$, which for Eq.~\eqref{app:eq:law} reads $-\varpi_{\rm eff}|H_h|^2/|I|^2$: amplification occurs if and only if $\omega<m(1-\alpha)\Omega_H$. The band edge measures $\alpha$ directly.

The solver integrates the Frobenius solutions from $x=0.08$ outward and decomposes the field at $r\ge60$ (or $30/\omega$) on the asymptotic in/out series; the same code, with the outgoing solution continued along a complex ray, gives the quasinormal frequencies. It reproduces Leaver's $m=1$--$4$ values to $2\times10^{-4}$, the Cardoso--Lemos--Yoshida co- and counter-rotating frequencies of Table~\ref{app:tab:rot} to the quoted digits, and satisfies the flux identity to $3\times10^{-5}$ in every scattering run. Figure~\ref{app:fig:law} collects the results.

\emph{Absorption frame.} The band edge follows $m(1-\alpha)\Omega_H$ to the resolution of the frequency grid, and the peak gain of $m=1$ falls from $5.1\%$ at $\alpha=0$ to $2.1$, $0.6$ and $0.1\%$ at $\alpha=0.25$, $0.5$ and $0.75$; at $\alpha=1$ both senses of rotation are absorbed at every frequency [Fig.~\ref{app:fig:law}(a,b)]. The $m=2$ and $m=3$ spectra have the same shape.

\emph{Reflectivity.} With the reflective horizon of Sec.~\ref{app:sec:qnmx}, $b/a=R_h=0.30\,e^{i130^\circ}$ at $\alpha=0$, the $m=1$ peak gain is $0.5$, $3.5$, $6.5$ and $10.5\%$ at $B/A=0.19$, $0.39$, $0.5$ and $0.61$, against $0.4$, $2.7$, $5.1$ and $8.1\%$ for the ideal horizon, in the same band. Interference with the centrifugal barrier makes the gain depend on the phase of $R_h$ ($2.7$--$9.5\%$ at the fitted modulus), but a partially reflecting horizon does not remove amplification.

\emph{Quasinormal modes.} Running the same law through the quasinormal problem at $B/A=0.40$ shifts the real parts of the co- and counter-rotating $m=2,3,4$ frequencies within the agreement of the measured frequencies with Cardoso--Lemos--Yoshida for $\alpha\le0.75$ [Fig.~\ref{app:fig:law}(c), Table~\ref{app:tab:rot}], so the ringdown does not fix $\alpha$. The frame parameter acts instead on the damping, as a co/counter asymmetry: at $\alpha=1$ co-rotating modes are under-damped by $5$--$14\%$ and counter-rotating $m=-2$ and $-3$ over-damped by $17$ and $34\%$, while $m=-4$ is nearly unchanged. The measured asymmetry (Table~\ref{app:tab:rot}) is opposite and larger, co-rotating modes over-damped and counter-rotating under-damped: the lattice horizon is not a rotating absorber of this kind.


\begin{thebibliography}{99}
\bibitem{Unruh1981} W.~G.~Unruh,
\newblock Phys. Rev. Lett. \textbf{46}, 1351 (1981).

\bibitem{Visser1998} M.~Visser,
\newblock Class. Quantum Grav. \textbf{15}, 1767 (1998).

\bibitem{BarceloLR} C.~Barcel\'o, S.~Liberati, and M.~Visser,
\newblock Living Rev. Relativity \textbf{14}, 3 (2011).

\bibitem{Schutzhold1998} R.~Sch\"utzhold and W. G. Unruh,
\newblock Phys. Rev. D \textbf{66}, 044019 (2002).

\bibitem{Visser2005} M.~Visser and S.~Weinfurtner,
\newblock Class. Quantum Grav. \textbf{22}, 2493 (2005).

\bibitem{Weinfurtner2011} S.~Weinfurtner, E.~W.~Tedford, M.~C.~J.~Penrice,
W.~G.~Unruh, and G.~A.~Lawrence,
\newblock Phys. Rev. Lett. \textbf{106}, 021302 (2011).

\bibitem{Torres2017} T.~Torres, S.~Patrick, A.~Coutant, M.~Richartz,
E.~W.~Tedford, S.~Weinfurtner,
\newblock Nat. Phys. \textbf{13}, 833 (2017).

\bibitem{Steinhauer2016} J.~Steinhauer,
\newblock Nat. Phys. \textbf{12}, 959 (2016).

\bibitem{Drori2019} J.~Drori, Y.~Rosenberg, D.~Bermudez, Y.~Silberberg,
U.~Leonhardt,
\newblock Phys. Rev. Lett. \textbf{122}, 010404 (2019).

\bibitem{Succi2001} S.~Succi,
\newblock \emph{The Lattice Boltzmann Equation for Fluid Dynamics and Beyond}
(Oxford University Press, 2001).

\bibitem{Chopard1998}
B. Chopard, M. Droz, \newblock \emph{Cellular Automata Modeling of Physical Systems}, (Cambridge University Press, 1998).

\bibitem{FHP1986} U.~Frisch, B.~Hasslacher, and Y.~Pomeau,
\newblock Phys. Rev. Lett. \textbf{56}, 1505 (1986).

\bibitem{CLY2004} V.~Cardoso, J.~P.~S.~Lemos, and S.~Yoshida,
\newblock Phys. Rev. D \textbf{70}, 124032 (2004).

\bibitem{Wolfram1983} S.~Wolfram,
\newblock Rev. Mod. Phys. \textbf{55}, 601 (1983).

\bibitem{LGCA}
D.H. Rothman, S. Zaleski, \newblock \emph{Lattice Gas Cellular Automata}, (Cambridge University Press, 2009).

\bibitem{Toffoli} T.~Toffoli, N.~Margolus,
\newblock \emph{Cellular Automata Machines}
(MIT Press, 1985).

\bibitem{CorleyJacobson1998} S.~Corley and T.~Jacobson,
\newblock Phys. Rev. D \textbf{57}, 6269 (1998).

\bibitem{JacobsonMattingly2000} T.~Jacobson and D.~Mattingly,
\newblock Phys. Rev. D \textbf{61}, 024017 (2000).

\bibitem{DiMolfetta2013} G.~Di~Molfetta, M.~Brachet, and F.~Debbasch,
\newblock Phys. Rev. A \textbf{88}, 042301 (2013).

\bibitem{Mendoza2010} M.~Mendoza, B.~M.~Boghosian, H.~J.~Herrmann, and S.~Succi,
\newblock Phys. Rev. Lett. \textbf{105}, 014502 (2010).

\bibitem{Mendoza2013} M.~Mendoza, S.~Succi, and H.~J.~Herrmann,
\newblock Sci. Rep. \textbf{3}, 3106 (2013).

\bibitem{tHooft2016} G.~'t~Hooft,
\newblock \emph{The Cellular Automaton Interpretation of Quantum Mechanics},
Fundamental Theories of Physics Vol.~185 (Springer, 2016).

\bibitem{Lallemand2000} P.~Lallemand, L.-S.~Luo,
\newblock Phys. Rev. E \textbf{61}, 6546 (2000).

\bibitem{Ginzburg2008} I.~Ginzburg, F.~Verhaeghe, and D.~d'Humi\`eres,
\newblock Commun. Comput. Phys. \textbf{3}, 427 (2008).

\bibitem{Guo2002} Z.~Guo, C.~Zheng, B.~Shi,
Phys. Rev. E \textbf{65}, 046308 (2002).

\bibitem{TorresJFM2018} T.~Torres, A.~Coutant, S.~Dolan,
S.~Weinfurtner, J. Fluid Mech. \textbf{857}, 291 (2018).

\bibitem{Leaver1985} E.~W.~Leaver,
\newblock Proc. R. Soc. Lond. A \textbf{402}, 285 (1985).

\bibitem{BCL2004} E.~Berti, V.~Cardoso, and J.~P.~S.~Lemos,
\newblock Phys. Rev. D \textbf{70}, 124006 (2004).

\bibitem{BasakMajumdar2003} S.~Basak and P.~Majumdar,
\newblock Class. Quantum Grav. \textbf{20}, 3907 (2003).

\bibitem{Zeldovich1971} Ya.~B.~Zel'dovich,
\newblock JETP Lett. \textbf{14}, 180 (1971).

\bibitem{Zeldovich1972} Ya.~B.~Zel'dovich,
\newblock Sov. Phys. JETP \textbf{35}, 1085 (1972).

\bibitem{Cromb2020} M.~Cromb, G.~M.~Gibson, E.~Toninelli, M.~J.~Padgett,
E.~M.~Wright, and D.~Faccio,
\newblock Nat. Phys. \textbf{16}, 1069 (2020).

\bibitem{Torres2020} T.~Torres, S.~Patrick, M.~Richartz,
S.~Weinfurtner,
\newblock Phys. Rev. Lett. \textbf{125}, 011301 (2020).

\bibitem{Unruh1995} W.~G.~Unruh, Phys. Rev. D \textbf{51}, 2827 (1995).

\bibitem{BroutMassar1995} R.~Brout, S.~Massar, R.~Parentani, and P.~Spindel, Phys. Rev. D \textbf{52}, 4559 (1995).

\bibitem{CorleyJacobson1996} S.~Corley and T.~Jacobson, Phys. Rev. D \textbf{54}, 1568 (1996).

\bibitem{UnruhSchutzhold2005} W.~G.~Unruh and R.~Sch\"utzhold, Phys. Rev. D \textbf{71}, 024028 (2005).
\end{thebibliography}
\end{document}